\documentclass[12pt,preprint]{aastex}
\usepackage{natbib}
\usepackage{subfigure}
\usepackage{float}

\def\msun{{\rm\,M_\odot}}

\def\msun{{\rm\,M_\odot}}

\def\h2{${\rm\,H_2}$}

\makeatletter 

\def\msun{{\rm\,M_\odot}}

\def\vol#1  {{{#1}{\rm,}\ }}

\newcount\refno
\newcount\rfno
\def\eq{$^{\the\refno\ }$\advance\refno by 1}
\def\ad{\advance\rfno by 1}

\def\clock{\count0=\time \divide\count0 by 60
     \count1=\count0 \multiply\count1 by -60 \advance\count1 by \time
     \number\count0:\ifnum\count1<10{0\number\count1}\else\number\count1\fi}

\def\myputfigure#1#2#3#4#5%
{\hskip0.03\textwidth\vskip#5pt
\makebox[0pt]{\hskip#2in
\includegraphics[width=#3\textwidth]{#1}}\vskip#4pt\hfill}

\newcount\refno
\newcount\rfno
\def\eq{$^{\the\refno\ }$\advance\refno by 1}
\def\ad{\advance\rfno by 1}

\begin{document}

\title{Far-Infrared Properties of Lyman Break Galaxies from Cosmological Simulations}
 
\author{
Renyue Cen$^{1}$
} 
 
\footnotetext[1]{Princeton University Observatory, Princeton, NJ 08544;
 cen@astro.princeton.edu}

\begin{abstract}

Utilizing state-of-the-art, adaptive mesh-refinement 
cosmological hydrodynamic simulations 
with ultra-high resolution ($114h^{-1}$pc) and large sample size ($\ge 3300$ galaxies of stellar mass $\ge 10^9\msun$),
we show how the stellar light of Lyman Break Galaxies at $z=2$ is distributed between optical/ultra-violet (UV) 
and far-infrared (FIR) bands.  
With a single scalar parameter for dust obscuration
we can simultaneously reproduce the observed UV luminosity function for the entire range ($3-100\msun$yr$^{-1}$)
and extant FIR luminosity function at the bright end ($\ge 20\msun$yr$^{-1}$). 
We quantify that galaxies more massive or having higher SFR
tend to have larger amounts of dust obscuration mostly due to a trend in column density 
and in a minor part due to a mass (or SFR)-metallicity relation.
It is predicted that the FIR luminosity function in the range ${\rm SFR}=1-100\msun$yr$^{-1}$ 
is a powerlaw with a slope about $-1.7$.
We further predict that there is a ``galaxy desert"
at ${\rm SFR_{\rm FIR}} < 0.02 ({\rm SFR_{UV}}/10 \msun {\rm yr}^{-1})^{2.1} \msun {\rm yr}^{-1}$
in the ${\rm SFR_{UV}}-{\rm SFR_{\rm FIR}}$ plane.
Detailed distributions of ${\rm SFR_{FIR}}$ at a fixed ${\rm SFR_{UV}}$ are presented.
Upcoming observations by ALMA should test this model.
If confirmed, it validates the predictions of the standard cold dark matter model and
has important implications on the intrinsic SFR function of galaxies at high redshift.

\end{abstract}
 
 
\section{Introduction}

The precise relation between optical/UV light detected and dust emission in the far infrared (FIR) of 
Lyman Break Galaxies \citep[LBGs;][]{2003Steidel} is difficult to establish observationally,
because of the faintness of the expected FIR luminosity \citep[e.g.,][]{1999Ouchi, 2000Adelberger}. 
In this work we study this relation using direct simulations of galaxy formation in the 
standard cosmological constant-dominated cold dark matter model \citep[LCDM; ][]{2010Komatsu}
in light of the capabilities of the upcoming Atacama Large Millimeter Array (ALMA) mission.
The outline of this paper is as follows.
In \S 2 we detail our simulations, method of making galaxy catalogs and a dust obscuration analysis method. 
Results are presented in \S 3, followed by conclusions given in \S 4.

\section{Simulations}\label{sec: sims}

\subsection{Hydrocode and Simulation Parameters}

We perform cosmological simulations with the adaptive mesh refinement (AMR) 
Eulerian hydro code, Enzo 
\citep[][]{1999bBryan, 2009Joung}.  
First we ran a low resolution simulation with a periodic box of $120~h^{-1}$Mpc on a side.
We identified a region centered on a cluster of mass of $\sim 2\times 10^{14}\msun$ at $z=0$
and then resimulate it with high resolution,
embedded in the outer $120h^{-1}$Mpc box.
The refined region for ``C" run has a size of $21\times 24\times 20h^{-3}$Mpc$^3$
and represents $1.8\sigma$ fluctuation on that volume.
The dark matter particle mass in the refined region is $1.3\times 10^7h^{-1}\msun$.
The refined region is surrounded by three layers (each of $\sim 1h^{-1}$Mpc) of buffer zones with 
particle masses successively larger by a factor of $8$ for each layer, 
which then connects with
the outer root grid that has a dark matter particle mass $8^4$ times that in the refined region.
We choose the mesh refinement criterion such that the resolution is 
always better than $114h^{-1}$pc physical, corresponding to a maximum mesh refinement level of $13$ at $z=0$.
The simulations include a metagalactic UV background
\citep[][]{1996Haardt}, a model for shielding of UV radiation by neutral hydrogen 
\citep[][]{2005Cen},  metallicity-dependent radiative cooling \citep[][]{1995Cen}
extended down to $10~$K \citep[][]{1972Dalgarno}
and all relevant gas chemistry chains for molecular hydrogen formation \citep[][]{1997Abel},
including molecular formation on dust grains \citep[][]{2009Joung}.
Star particles are created in cells that satisfy a set of criteria for 
star formation proposed by \citet[][]{1992CenOstriker}.
%
Supernova feedback from star formation is modeled following \citet[][]{2005Cen}.
We allow the entire feedback processes to be hydrodynamically coupled to surroundings
and subject to relevant physical processes, such as cooling and heating. 
%
See \citet[][]{2010Cen} for all other simulation details and physical treatments.
We use the following cosmological parameters that are consistent with 
the WMAP7-normalized \citep[][]{2010Komatsu} LCDM model:
$\Omega_M=0.28$, $\Omega_b=0.046$, $\Omega_{\Lambda}=0.72$, $\sigma_8=0.82$,
$H_0=100 h {\rm km s}^{-1} {\rm Mpc}^{-1} = 70 {\rm km s}^{-1} {\rm Mpc}^{-1}$ and $n=0.96$.

\subsection{Simulated Galaxy Catalogs}

We identify galaxies in our high resolution simulations using the HOP algorithm 
\citep[][]{1999Eisenstein}, operated on the stellar particles, which is tested to be robust.
Satellites within a galaxy are clearly identified separately.
The luminosity of each stellar particle at each of the Sloan Digital Sky Survey (SDSS) five bands 
is computed using the GISSEL stellar synthesis code \citep[][]{Bruzual03}, 
by supplying the formation time, metallicity and stellar mass.
Collecting luminosity and other quantities of member stellar particles, gas cells and dark matter 
particles yields
the following physical parameters for each galaxy:
position, velocity, total mass, stellar mass, gas mass, 
mean formation time, 
mean stellar metallicity, mean gas metallicity,
star formation rate,
luminosities in five SDSS bands (and various colors) and others.
At a spatial resolution of $109$pc with nearly 5000 well resolved galaxies at $z=2$,
this simulated galaxy catalog presents an excellent tool to study galaxy formation and evolution.

\subsection{Modeling Dust Obscuration}

A fully self-consistent modeling would be difficult, given our lack of 
knowledge of the distribution of dust and its properties.
Here we take a simplified approach.
Given the 3-d distribution of gas with varying metallicity and stellar particles distributed within it,
the observed SFR at a rest-frame UV wavelength $\lambda$ for the galaxy is computed as  
\begin{equation}
\label{eq:SFRUV}
{\rm SFR}_{\rm UV, \lambda} = \sum_i {\rm sfr}_i (1-e^{-\tau_\lambda(\vec r_i\rightarrow {\rm obs})}),
\end{equation}
\noindent
where $\tau_\lambda(\vec r\rightarrow {\rm obs})$ is the extinction optical depth at some UV wavelength $\lambda$ for  
an individual stellar particle $i$ of star formation rate ${\rm sfr}_i$ in the galaxy 
from its individual location $\vec r_i$ to the observer:
\begin{equation}
\label{eq:taulambda}
\tau_\lambda(\vec r\rightarrow {\rm obs}) = (A_V^\prime/1.086) f \beta_\lambda {\bar Z_i}(\vec r\rightarrow {\rm obs}) 
N_{H,i}(\vec r\rightarrow {\rm obs}),
\end{equation}
\noindent
where $A_V^\prime = 5.3 \times 10^{-22}$ is visual extinction $A_V$ per unit hydrogen column density per unit solar metallicity 
for $R_V=3.1$ \citep[][]{2011Draine} and
$\beta_\lambda\equiv A_\lambda/A_V$ (a fitting function) is taken from \citet[][]{1989Cardelli};
${\bar Z_i}(\vec r\rightarrow {\rm obs})$ is the column density-weighted mean metallicity of gas obscuring 
the stellar particle $i$ in solar units 
and $N_{H,i}(\vec r\rightarrow {\rm obs})$ is the integrated hydrogen column density from the stellar particle $i$ to the observer.
Note that in Equation~(\ref{eq:SFRUV}) the calculation is based on 3-d distributions of stellar particles 
that each are subject to their own integrated optical depth and the sum is over all the memeber stellar particles,
typically of number $10^5-10^6$ for a galaxy of stellar mass $10^{11}\msun$.
In Equation~(\ref{eq:taulambda}) $f$ is a dimensionless parameter that we will adjust 
such that the simulated LBG UV luminosity function matches observations;
f should be of order unity, if dust properties for galaxies at $z\sim 2$ are not drastically different from those
derived locally
and our galaxy formation model is realistic.
As we will see below, the required value of $f$ is indeed close to unity with an adopted extinction law 
that is also close to those derived locally.
Thus, the dust extinction of SFR at a specific UV band 
is a good proxy of the overall extinction of SFR in the optical-to-UV regime.
We will use the $1700\AA$ band for subsequent analysis.
The portion of the SFR that does not escape in UV/optical is assumed to be converted to FIR SFR:
\begin{equation}
\label{eq:SFRFIR}
{\rm SFR}_{\rm FIR} = \sum_i {\rm sfr}_i - {\rm SFR}_{\rm UV, \lambda}.
\end{equation}
\noindent
For each galaxy we place 95 random observers in its sky at infinity for results presented in the next section.
This sampling is adequate and results are converged statistically.

\section{Results}

\begin{figure}[ht]
\centering
\vskip -0.5cm
\resizebox{4.5in}{!}{\includegraphics[angle=270]{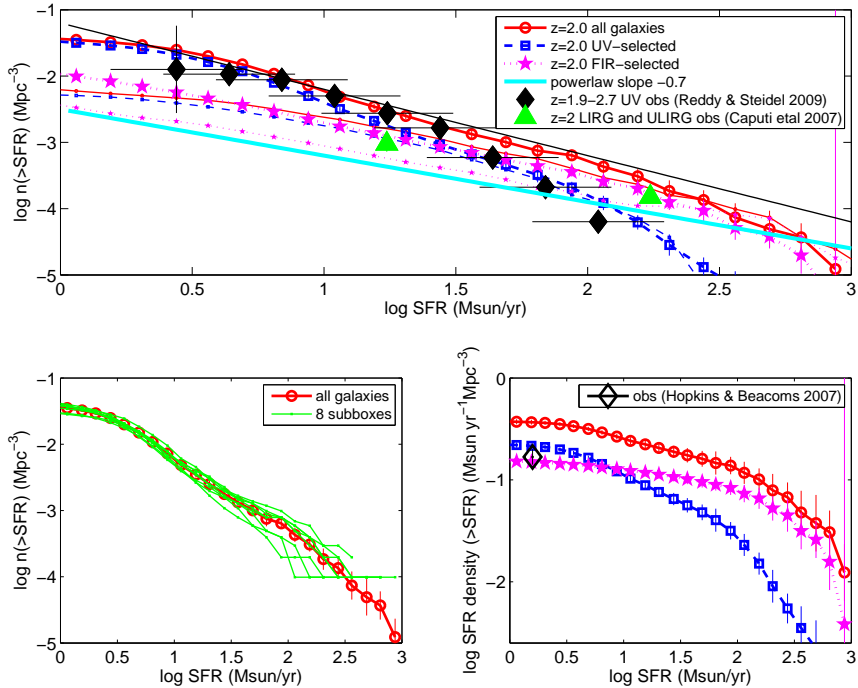}}
\vskip -0cm
\caption{\footnotesize 
{\it Top panel}: 
cumulative total SFR function at $z=2$ (red circles),
cumulative UV and FIR SFR functions in blue squares and magenta stars, respectively.
Black diamonds are LBG observations $z=1.9-2.7$ from \citet[][]{2009Reddy};
two green triangles are LIRG and ULIRG observational data from \citet[][]{2007Caputi}.
We convert to SFR of observational data from $M_{\rm AB}(1700\AA)$ using the standard
conversion formula, ${\rm SFR} = 6.1\times 10^{-[8+0.4 M_{\rm AB}(1700\AA)]}\msun$/yr
\citep[][]{1998Kennicutt} in the AB magnitude system \citep[][]{1974Oke}.
Solid magenta line indicates a powerlaw slope of $-0.7$ (corresponding to a slope of $-1.7$ 
for the differential ${\rm SFR_{FIR}}$ function).
Thin solid black line indicates a powerlaw slope of $-1$ (corresponding to a slope of $-2$ for the differential function).
The three thin curves of color red, blue and magenta, respectively, correspond their thick 
counterparts but from a lower resolution simulation with four times poorer spatial and eight times 
poorer mass resolutions.
{\it Bottom left panel}: 
the eight green curves represent the cumulative total SFR function in the eight octant volumes;
the average of the green curves is the redshift circles (also shown in the top panel).
{\it Bottom right panel}: 
Cumulative light densities for total (red circles), UV galaxies (blue squares) and
FIR galaxies (magenta stars), respectively, at $z=2$.
Also show as a black diamond is the observed data at $z=2$ compiled by 
\citet[][]{2006HopkinsA} with $1\sigma$ errorbar.
}
\label{fig:LFz2000}
\end{figure}

The top panel of Figure~\ref{fig:LFz2000} shows
the SFR functions for total SFR,  UV and FIR selected galaxies, respectively.
We have adjusted the parameter $f$ in Equation ~\ref{eq:taulambda} 
to be $f=1.4$ to arrive at the excellent match between the computed UV SFR function and 
the observations at $z\sim 2$.
We note that $f$ could be $1$ if one had 
adopted a slightly different $R_v$ \citep[][]{1989Cardelli}.
In any case, the results with $f=1$ with $R_v=3.1$ differ only slightly 
from the case with $f=1.4$ shown here and UV SFR function in that case is
consistent with the observations within the errorbars.
This also suggests that our overall results are robust and insensitive to small variations
of uncertain parameters for the dust model within a reasonable range.
It also implies that dust properties at $z\sim 2$ are not significantly different from those of local dust.

\begin{figure}[ht]
\centering
\vskip -0cm
\resizebox{5.0in}{!}{\includegraphics[angle=270]{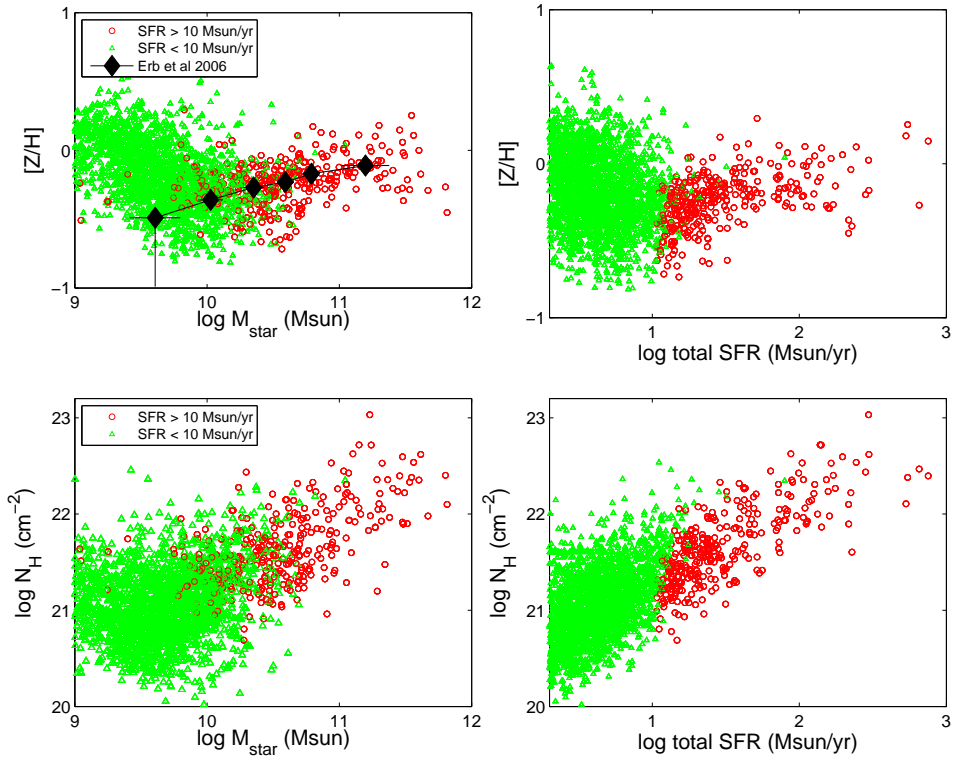}}
\vskip -0cm
\caption{\footnotesize 
{\it Top left panel}:
column density weighted gas metallicity averaged over the entire galaxy as a function of stellar mass.
Shown in black diamond is observations from \citet[][]{2006Erb} at $z\sim 2$.  
{\it Top right panel}: column density weighted gas metallicity averaged over the entire galaxy as a function of SFR.
{\it Bottom left panel}:
radially integrated total column density as a function of stellar mass.
{\it Bottom right panel}:
radially integrated total column density as a function of SFR.
In all the panels red symbols have SFR greater than $10\msun$/yr
and green symbols less than $10\msun$/yr.
}
\label{fig:comb}
\end{figure}

After matching the observed UV SFR function, we see that the predicted FIR SFR function agrees remarkably well 
with the observed LIRG and ULIRG data at $z=2$ (top panel).
As a consistency check, we show
in the bottom right panel of Figure~\ref{fig:LFz2000} 
the cumulative SFR density at $z=2$.
Here we see both the UV SFR density and FIR SFR density agree well with observations.
We see that, while the directly observed UV SFR density should 
be roughly equal to the directly observed FIR SFR density,
at face value, the UV SFR density is somewhat higher than FIR SFR density.
Our results suggest that
galaxies with higher SFR tend to have relatively larger obscuration in UV/optical than galaxies with lower SFR, resulting in
a steepening UV luminosity function at the luminous end. The underlying cause will be discussed in Figure~\ref{fig:comb}.

Our previous studies \citep[][]{2010Cen, 2011bCen}
indicate that the ``C" run used is positively biased over
the cosmic mean by a factor of $\sim 2$.
Taking that into account, we find that the simulated SFR function as well SFR density
becomes too low compared to observed ones.
A plausible adjustment is to the stellar IMF.
The results shown above uses an top-heavy IMF that 
produces twice the UV light output per unit SFR than the standard Salpeter function.
This provides intriguing evidence for top-heavy IMF at high redshift, consistent with other independent considerations
\citep[e.g.,][]{2005Baugh, 2008Dave, 2008vanDokkum}. 
The abundance of massive, rare objects is expected to depend on box size as well as the overdensity of the environment.
We assess this effect as follows.
We divide the simulation box into eight equal-volume octants and 
compute the SFR function for each of the eight octants.
The results for the eight SFR functions are shown as green curves in the bottom left panel of Figure~\ref{fig:LFz2000}.
We note two points here.
First, at ${\rm SFR} \le 30\msun$/yr the SFR function is very well converged and does not appear to sensitively depend on
environment.
Second, substantial variations are visible at ${\rm SFR}\ge 100\msun$/yr, which suggests
that the abundance of galaxies with SFR higher than $100\msun$/yr depends sensitively on density environment
and our positively biased simulation box likely has produced some over-abundance of galaxies with ${\rm SFR}\ge 100\msun$/yr
relative to galaxies with ${\rm SFR}\le 100\msun$/yr;
the computed UV SFR at ${\rm SFR}\ge 100\msun$/yr in this simulation
lies above the observed points is thus not inconsistent.

A comparison between the thin and thick curves in the top panel of 
Figure~\ref{fig:LFz2000} indicates that the resolutions achieved in the higher resolution run
is required in order to provide an adequate match to observations.
The lower (four times spatially and eight times in mass) resolution simulation of the same volume
suffers from the two shortcomings. 
First, there is a slight overproduction of the highest SFR ($\ge 200\msun$~yr$^{-1}$) galaxies 
in the lower resolution simulation, which is due to a combination of slight overmerging and
higher gas reservoir in the lower resolution run.
Second, there is a significant underproduction of lower SFR ($\le 200\msun$~yr$^{-1}$) galaxies 
in the lower resolution simulation due to lower resolution.
Taking into account these two effects, 
the results can be understood and our main predictions
on the faint slope and galaxy desert (see below) remain robust.

Our model makes several predictions.
The first is that the differential FIR SFR function 
displays a nearly perfect powerlaw of slope about $-1.7$ below ${\rm SFR_{\rm FIR}}\sim 100\msun$/yr at $z=2$.
We attribute this outcome to a combination of three physical factors:
(1) the intrinsic differential SFR function is steeper than than $-1.7$ but close to $-2$, as indicated
by the thin black line in the top panel of 
Figure~\ref{fig:LFz2000};
(2) on averagge, higher SFR galaxies have higher dust optical depth 
 (as discussed in detail in Figure~\ref{fig:comb} below) that tends to flatten the FIR SFR function;
(3) there is a significant dispersion of ${\rm SFR_{\rm FIR}}$ at a fixed intrinsic SFR 
(see Figure~\ref{fig:uvFIR} below) that also smoothes and flattens the FIR SFR function.
This predictions can be tested by ALMA observations, and if confirmed, will provide 
evidence that the intrinsic SFR function is close to a powerlaw with a slope that is steeper than $-1.7$
in the SFR range $10-300\msun$yr$^{-1}$. 
We attribute this behavior to a large dispersion 
of SFR at a fixed halo mass but will address it in more detail separately.

Given this slope, most of the FIR light is concentrated at the bright end.
In terms of cumulative galaxy number density 
we find that UV and FIR selected samples
are expected to have comparable abundances at ${\rm SFR}\ge 20-40\msun$/yr.
In terms of cumulative SFR density we find that FIR selected galaxies with ${\rm SFR}\ge 10\msun$/yr 
dominate over UV selected galaxies with ${\rm SFR}\ge 10\msun$/yr;
the reverse is true at ${\rm SFR}<10\msun$/yr.
Reading directly from simulations we find that FIR selected galaxies with FIR ${\rm SFR}\ge 10\msun$/yr
contain $78\%$ of total FIR light density,
whereas UV selected galaxies with UV ${\rm SFR}\ge 10\msun$/yr
contain only $50\%$ (the actual number may be still lower, since our simulations likely
have under-estimated the number density of galaxies below ${\rm SFR}\le 3\msun$/yr, 
below which a flattening of the UV SFR function is seen in the top panel of Figure~\ref{fig:LFz2000}). 
Note that while a Schechter function normally fits halo functions well,
it does not provide an adequate fit to the FIR SFR function, due to 
large dispersions of SFR at fixed halo masses mentioned above.
Our results suggest that the observed UV-selected LBGs detected at ${\rm SFR}\ge$ a few $\msun$/yr at $z=2-3$ 
can account for the bulk of the FIR background at $z\sim 2-3$,
consistent with earlier independent observational assessments \citep[e.g.,][]{1999Smail, 2000Adelberger, 2009Chapman}. 

Needless to say, our model implies that UV and FIR selected galaxies form a complementary pair of populations
that are drawn from the same underlying general galaxy population.
This point has been noted by others \citep[e.g.,][]{1998Sawicki, 1999Meurer, 2001Shapley, 2001Papovich, 2001Calzetti}.

\begin{figure}[ht]
\centering
\vskip -0.0cm
\resizebox{4.5in}{!}{\includegraphics[angle=270]{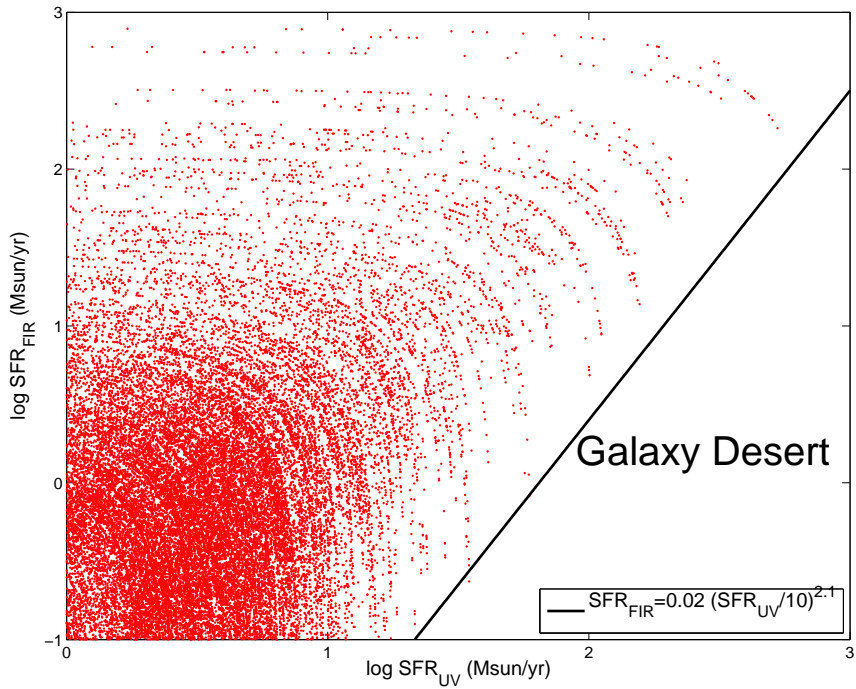}}
\vskip 0cm
\caption{\footnotesize 
each dot is a galaxy in the plane of  UV and FIR detected SFR at $z=2$.
The solid line is ${\rm SFR_{\rm FIR}} = 0.02 [{\rm SFR{UV}}/10 \msun {\rm yr}^{-1}]^{2.1} \msun {\rm yr}^{-1}$.
}
\label{fig:uvFIR}
\end{figure}

Figure~\ref{fig:uvFIR} shows a scatter plot of galaxies in the 
${\rm SFR_{\rm UV}}-{\rm SFR_{FIR}}$ plane.
We see a nearly complete empty space at the lower right corner of the plot,
with ${\rm SFR_{\rm FIR}} < 0.02 [{\rm SFR{UV}}/10 \msun {\rm yr}^{-1}]^{2.1} \msun {\rm yr}^{-1}$,
which we shall call the ``galaxy desert".
The physical reason for this nearly complete absence of galaxies with high UV SFR and low FIR SFR rate is that
the dust optical depth of galaxies increases with SFR.
This second prediction of our model should be testable by ALMA observations.

\begin{figure}[ht]
\centering
\vskip -0cm
\resizebox{5.0in}{!}{\includegraphics[angle=270]{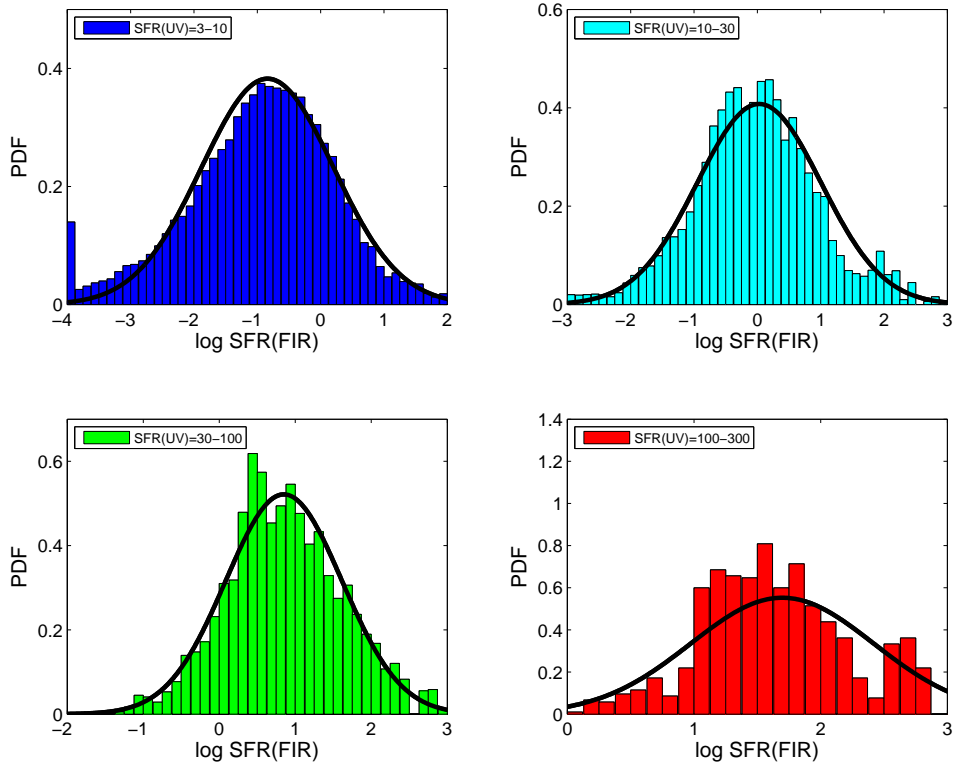}}
\vskip -0cm
\caption{\footnotesize 
shows four distributions of FIR SFR for LBG galaxies at each of the four 
UV SFR values of
$3-10\msun$/yr (top left),
$10-30\msun$/yr (top right),
$30-100\msun$/yr (bottom left)
and 
$100-300\msun$/yr (bottom right), respectively,
at $z=2$.
Each black curve is a gaussian fit with its parameters listed in Table 1.
}
\label{fig:firSFR}
\end{figure}

\begin{deluxetable}{ccc}
\tablecolumns{3}
\tablewidth{0pc}
\tablecaption{Parameters for gaussians in 
Figure~\ref{fig:firSFR} 
with $\log {\rm SFR}_{\rm UV}$ being the variable
\label{tab:table1}}
\tablehead{
\colhead{${\rm SFR}_{\rm UV}(\msun$/yr)} & \colhead{mean} & \colhead{dispersion}}
\startdata
3-10  & -0.84 & 1.0  \\
10-30  & 0.030 & 0.98  \\
30-100  & 0.85 & 0.76  \\
100-300  & 1.7 & 0.72  
\enddata
\end{deluxetable}

Figure~\ref{fig:firSFR} 
dissects the information contained in Figure~\ref{fig:uvFIR} further 
and shows a set of distributions of ${\rm SFR_{FIR}}$ at a given range of ${\rm SFR_{UV}}$.
We see that for LBGs with ${\rm SFR}_{\rm UV}=10-100\msun$/yr, the distributions
are well fitted by gaussians (the black curves) using $\log {\rm SFR}_{\rm UV}$ as the variable.
In the lowest ${\rm SFR_{UV}}$ (${\rm SFR}_{\rm UV}=3-10\msun$/yr)
we see a slight tendency of the ${\rm SFR_{FIR}}$ distribution to skew towards the low ${\rm SFR_{FIR}}$ end,
indicative of increasingly diminishing dust obscuration for galaxies with low SFR.
In the highest ${\rm SFR_{UV}}$ (${\rm SFR}_{\rm UV}=100-300\msun$/yr),
the ${\rm SFR_{FIR}}$ distribution is significantly skewed to the high ${\rm SFR_{FIR}}$ end
for the same physical reason.
We list the parameters of the best gaussian fit of 
${\rm SFR_{FIR}}$ distributions for all ${\rm SFR_{UV}}$ bins in Table 1.
These predictions should be verifiable by ALMA observations.

Finally, we examine the underlying cause of the generally differential rate
of dust obscuration seen in prior figures where higher SFR galaxies are more dust obscured.
Figure~\ref{fig:comb} shows gas metallicity and gas column density as a function
of stellar mass and total SFR, respectively.
Examination of the top left panel of 
Figure~\ref{fig:comb} indicates that 
in the stellar mass range $M_{\rm star}\ge 10^{10}\msun$ 
there is a positive correlation between gas metallicity and stellar mass,
in agreement with the observed, so-called mass-metallicity relation at $z\sim 2$ of 
\citet[][]{2006Erb}.
While this is not the focus of our study here,
the agreement is quite remarkable but consistent with 
the agreement that is found between simulations and observations
with respect to the metallicity distribution of damped Lyman alpha systems
in an earlier study \citep[][]{2010Cen}.
A comparison between the two top and two bottom panels of Figure~\ref{fig:comb} 
clearly indicates that the correlation between column density and stellar mass or SFR
is about three times stronger than that between gas metallicity and stellar mass or SFR.
This suggests that the general trend of larger dust obscuration for larger
stellar mass or SFR is mostly due to a trend in column density in the same sense,
but positively aided by a mass (or SFR)-metallicity trend.
This overall trend gives an integral constraint on the total optical depth.
The actual distribution of dust optical depth at a given galaxy mass (or SFR)
or even for a given galaxy viewed at different angels
has large dispersions due to the clumpy distribution of gas with varying metallicity,
resulting in a wide FIR SFR distribution within a narrow UV SFR range, as quantified
in Figure~\ref{fig:firSFR}.

\section{Conclusions}

Using state-of-the-art, adaptive mesh-refinement Eulerian cosmological hydrodynamic simulations 
with high resolution ($114h^{-1}$pc), large sample size ($\ge 3300$ galaxies of stellar mass $\ge 10^9\msun$) 
and a physically sound treatment of relevant processes,
we examine the properties of LBGs at $z=2$ with respect to their partitioning of UV and FIR light.  
Using a single scalar parameter 
that relates the amount of dust obscuration to the product of hydrogen column
density and gas metallicity to model dust obscuration along each line of sight (i.e., a dust extinction law 
derived in our local universe),
we find that the observed UV luminosity function for the entire range and FIR luminosity function at the bright end
can be simultaneously reproduced.
Our theoretical modeling affirms the aesthetically appealing picture where
UV and FIR selected galaxies at $z\sim 2$ are drawn from the same general galaxy population.
The observationally different manifestations are merely due to the known fact 
that each galaxy is seen through its own unique set of dust screens at a given viewing angle.
Star forming galaxies that are more massive or have higher SFR
tend to have larger amounts of dust obscuration at high redshift.

We predict that the FIR luminosity function in the range ${\rm SFR}=1-100\msun$/yr 
is a powerlaw with a slope $-1.7$ with uncertainty of $\sim 0.1$.
We further predict that there is a ``galaxy desert"
at ${\rm SFR_{\rm FIR}} < 0.02 ({\rm SFR_{UV}}/10 \msun {\rm yr}^{-1})^{2.1} \msun {\rm yr}^{-1}$
in the ${\rm SFR_{UV}}-{\rm SFR_{\rm FIR}}$ plane.
Detailed distributions of ${\rm SFR_{FIR}}$ at an observed ${\rm SFR_{UV}}$ are quantified 
and can be used to further test the model.
We expect that upcoming observations by ALMA should be able to test these predictions
hence ultimately the standard cosmological model with respect to its properties on sub-megaparsec scales.
If ALMA observations confirms the predicted faint end slope of the FIR luminosity function,
it would imply that the intrinsic SFR function of galaxies may be closer to a powerlaw of a slope
at least as steep as $-2$ in the range ${\rm SFR}=3-100\msun$ than a Schechter function.


\vskip 1cm

I would like to thank 
Computing resources were in part provided by the NASA High-
End Computing (HEC) Program through the NASA Advanced
Supercomputing (NAS) Division at Ames Research Center.
I thank an anonymous referee for a constructive report.
This work is supported in part by grants NAS8-03060 and NNX11AI23G.


\end{document}